\documentclass{article}
\usepackage{graphicx} 
\usepackage{xcolor}
\usepackage{amsfonts,amsmath,amssymb}
\usepackage{url}
\usepackage{hyperref}
\usepackage{bbm}

\usepackage{geometry}
\DeclareFontFamily{OT1}{cmtt}{\hyphenchar \font=-1}
\DeclareFontFamily{\encodingdefault}{\ttdefault}{\hyphenchar\font=`\-}
\DeclareFontFamily{T1}{cmtt}{\hyphenchar \font=45}

\title{Historical and Multichain Storage Proofs
}

\author{
Marek Kirejczyk \\
vlayer Labs \\
\texttt{marek@vlayer.xyz} \\ 
\and
Maciej Kalka\footnote{Corresponing author} \\
vlayer Labs \\
\texttt{maciej@vlayer.xyz} \\
\and 
Leonid Logvinov \\
vlayer Labs \\
\texttt{leon@vlayer.xyz} \\ \\
}
\date{November 2024}

\begin{document}

\maketitle
\begin{abstract}
This paper presents a comprehensive analysis of storage proofs in the Ethereum ecosystem, examining their role in addressing historical and cross-chain state access challenges. 
We systematically review existing approaches to historical state verification, comparing Merkle Mountain Range (MMR) and Merkle-Patricia trie (MPT) architectures. 
An analysis involves their respective performance characteristics within zero-knowledge contexts, where performance challenges related to Keccak-256 are explored. 
The paper also examines the cross-chain verification, particularly focusing on the interactions between Ethereum and Layer 2 networks. 
Through careful analysis of storage proof patterns across different network configurations, we identify and formalize three architectures for cross-chain verification. 
By organizing this complex technical landscape, this analysis provides a structured framework for understanding storage proof implementations in the Ethereum ecosystem, offering insights into their practical applications and limitations.
\end{abstract}

\newpage
\tableofcontents

\newpage
\section{Introduction}

The Ethereum blockchain and the broader Ethereum Virtual Machine (EVM) landscape have emerged as the dominant blockchain ecosystem, significantly due to their network effects, which manifest in various forms such as a loyal user base, a thriving developer community, and substantial financial liquidity~\cite{Khan2022}. 
As the first smart contract platform to achieve widespread adoption, Ethereum has established itself as a cornerstone of decentralized applications (dApps), creating an environment where both users and developers are incentivized to engage, innovate, and invest.

Central to Ethereum's success is the EVM, a virtual machine that facilitates the execution of smart contracts. 
One of the unique features of EVM-based smart contracts is their unprecedented access to the "world state", enabling all smart contracts on the same chain to interact with each other. 
This capability has been instrumental in creating a highly interconnected ecosystem where composability -- the ability to combine various protocols and applications -- drives continuous innovation and complex financial instruments such as decentralized finance products (i.e. money lego)~\cite{Buterin2013}.
Despite these strengths, the EVM faces two significant limitations that constrain its utility and expressiveness: access to historical state and the state of other chains within the ecosystem. 
Historical state access refers to the ability to query past states of the blockchain, while multichain state access is essential for interoperability between different blockchain networks, particularly between Ethereum and its Layer 2 solutions~\cite{gangwal2022}.

We investigate storage proofs as a comprehensive solution to these limitations.
Storage proofs enable verifiable access to the blockchain states by providing cryptographic evidence of data consistency and integrity. 
However, while storage proofs represent a significant step forward, they come with their own set of challenges, particularly in terms of performance and developer experience. 
The complexity of implementing and verifying storage proofs can deter developers while the performance overhead can reduce the efficiency of smart contracts, impacting the overall user experience.

Our analysis presents two distinct approaches for historical state verification using Merkle Mountain Range (MMR) and Merkle-Patricia trie (MPT) structures. 
We demonstrate that while MMR provides efficient proof generation, MPT offers superior flexibility for managing historical data at any depth.
Furthermore, we formalize three distinct patterns for cross-chain verification: L2$\rightarrow$L1, L1$\rightarrow$L2, and L2$\rightarrow$L2, accounting for the asymmetric security relationships between layers and their varying finality characteristics.
Additionally, the paper addresses performance challenges related to using Keccak-256 in zero-knowledge contexts, and analyzes alternative ZK-friendly hash functions. 

The paper is structured as follows: Section 2 provides the preliminaries, covering foundational data structures such as Merkle trees, Patricia tries, Merkle-Patricia tries, and their implementation in Ethereum. 
This section also introduces versioned data structures and conceptualizes Ethereum as a data structure. 
Section 3 examines the mechanics of storage proofs in detail, progressing from basic storage proofs and hierarchical proofs to an analysis of historical state verification using MMR and MPT approaches. 
The section concludes with a formalization of three architectures for multichain verification.
Finally, Section 4 summarizes the findings and discusses their implications for the broader Ethereum ecosystem.

\section{Preliminaries}

This section explores data structures used in Ethereum.
We begin with Merkle trees and Merkle proofs -- the fundamental building blocks for verifying data integrity in blockchains.
Understanding Merkle trees and proofs leads us to Patricia tries, which Ethereum uses for efficient key-value storage.
These two structures combine into Merkle-Patricia tries, giving Ethereum both efficient storage and cryptographic verification capabilities.
Building on this foundation, we explore the Ethereum Merkle-Patricia trie, an essential component of the Ethereum blockchain.
We then look at how these tries are versioned in Ethereum to track state changes over time.
Finally, we examine how these components fit together to form Ethereum's data architecture, setting the stage for a secure and scalable blockchain ecosystem.

\subsection{Merkle tree}

A Merkle tree (or hash tree) is a data structure used to efficiently and securely verify the integrity of data.
It was conceptualized by Ralph Merkle in 1980 as a solution for verifying contents of large datasets~\cite{10.1007/3-540-48184-2_32}.
The structure is a binary tree with three types of nodes:

\begin{itemize}
\item Leaf Nodes: Each contains a hash of a data block
\item Non-Leaf Nodes: Each contains a hash of the concatenation of its child nodes' hashes
\item Root Hash: The single hash at the top of the tree, representing the entire dataset
\end{itemize}

By comparing just the root hashes, large datasets can be verified for consistency. 
The process of verifying a particular data block involves computing the hash of the block and comparing it up the tree, which requires logarithmic time in terms of the number of blocks.
Due to logarithmic time complexity for insertions and deletions Merkle trees are suitable for large-scale data structures.
Moreover, use of cryptographic hashes ensures data integrity, making it difficult to alter the data without detection. Even a slight change in the data would result in a completely different hash.

\subsection{Merkle proof}
A Merkle proof is a method used to verify that a particular data block is part of a larger dataset without verifying the entire dataset. 
It leverages the structure and properties of Merkle trees to achieve efficient and secure verification.
To build a Merkle proof for given data block in a Merkle tree we use following components
\begin{itemize}
\item {Target Hash:} The hash of the data block that needs to be verified.
\item {Sibling Hashes:} The hashes of the sibling nodes along the path from the target hash to the root hash. These are required to reconstruct the path in the Merkle tree.
\item {Root Hash:} The hash at the top of the tree, which represents the entire dataset.
\end{itemize}

The verification process starts with computation of hash of target data block.
Then sequentially combine the target hash with the sibling hashes, computing the hash of each concatenated pair, to reconstruct the path up to the root hash.
The final computed hash is compared with the provided root hash. If they match, the target block is verified as part of the dataset.

\begin{figure}[ht!]
    \centering
    \includegraphics[width=\textwidth]{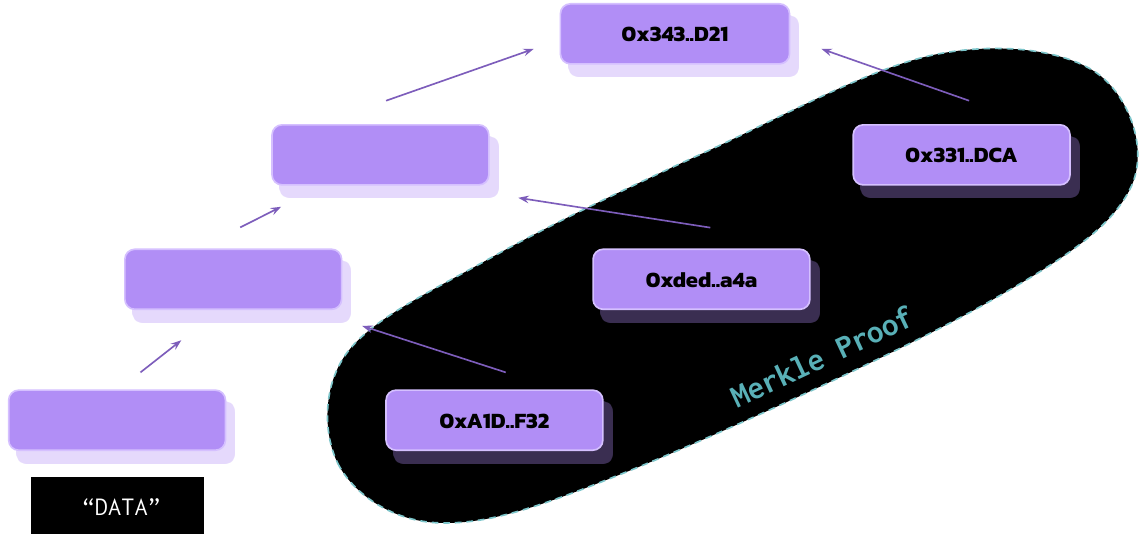}
    \caption{The Merkle proof (path) in a Merkle tree}
    \label{fig:merkle-proof}
\end{figure}

Figure~\ref{fig:merkle-proof} illustrates the concept of a Merkle proof within a Merkle tree. 
The diagram depicts a specific path from a target data block to the root hash, showcasing how a Merkle proof verifies the inclusion of a specific data block within a larger dataset.
Arrows indicate the combination of sibling hashes along the path, ultimately leading to the root hash.

The efficiency of a Merkle proof lies in its logarithmic complexity relative to the size of the dataset. 
For a dataset with $n$ blocks, the proof requires $\log(n)$ sibling hashes. 
This makes verification fast and scalable, especially for very large datasets.
Moreover, security of a Merkle proof is ensured by the cryptographic hashes used in the Merkle tree. 
Any alteration in the data would change the corresponding leaf hash, and subsequently the root hash, allowing for detection of tampered data.

\subsection{Patricia (Radix) trie}
A Patricia trie (Practical Algorithm to Retrieve Information Coded in Alphanumeric), or Radix trie, is a compressed version of a trie data structure that is used to store a set of strings. 
Invented by Donald R. Morrison in 1968~\cite{10.1145/321479.321481}, it optimizes trie structures by combining nodes that have a single child, thus reducing the space complexity.
\begin{figure}[ht!]
    \centering
    \includegraphics[width=\textwidth]{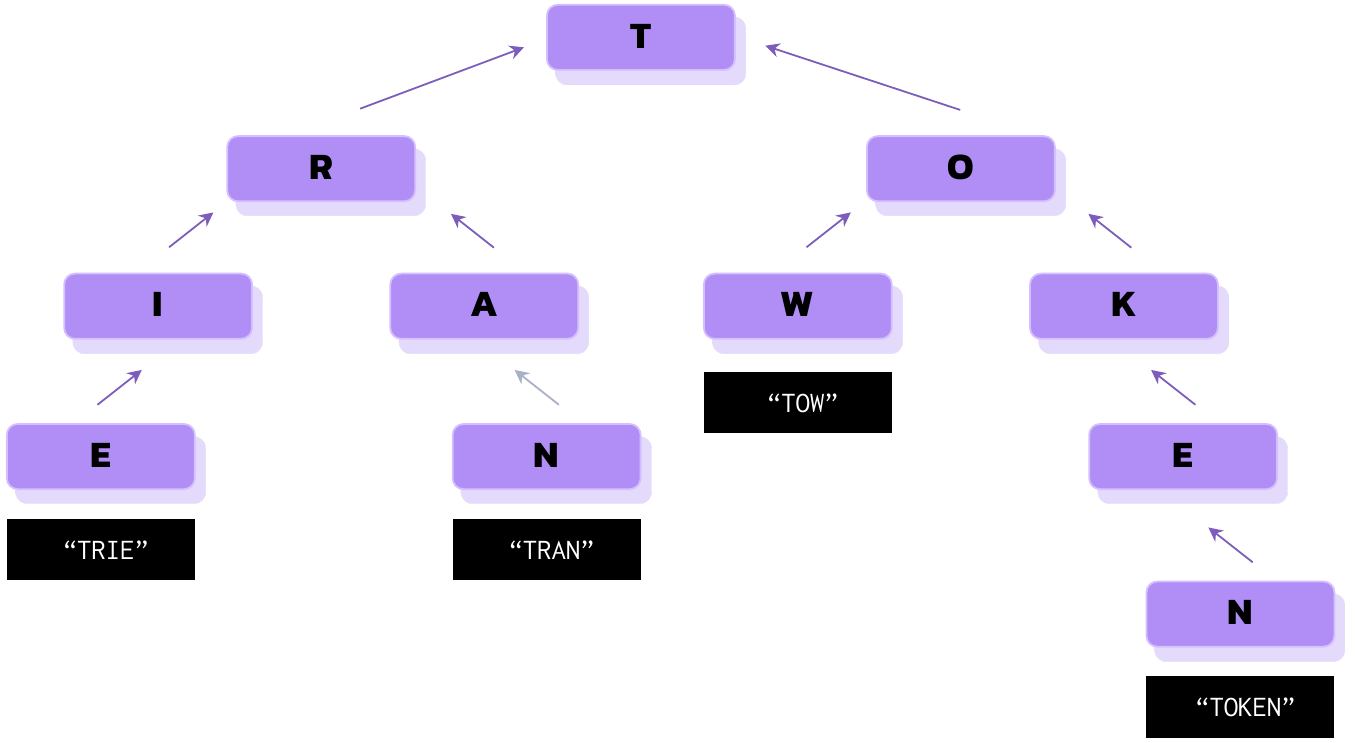}
    \caption{The Patricia trie}
    \label{fig:patricia-trie}
\end{figure}
\noindent
In the Fig.~\ref{fig:patricia-trie} an example of Patricia trie is shown. 
The main feature of the Patricia trie is that it provides an optimized $O(k)$ search time, where $k$ is the length of the search key, making it faster for searching than other tree-based data structures.
Nodes with a single child are merged, reducing the height of the tree, which leads to more efficient storage and retrieval operations.
Due to the merging of single-child nodes, it uses space more efficiently than a simple trie.

\subsection{Merkle-Patricia trie}
\begin{figure}[ht!]
    \centering
    \includegraphics[width=\textwidth]{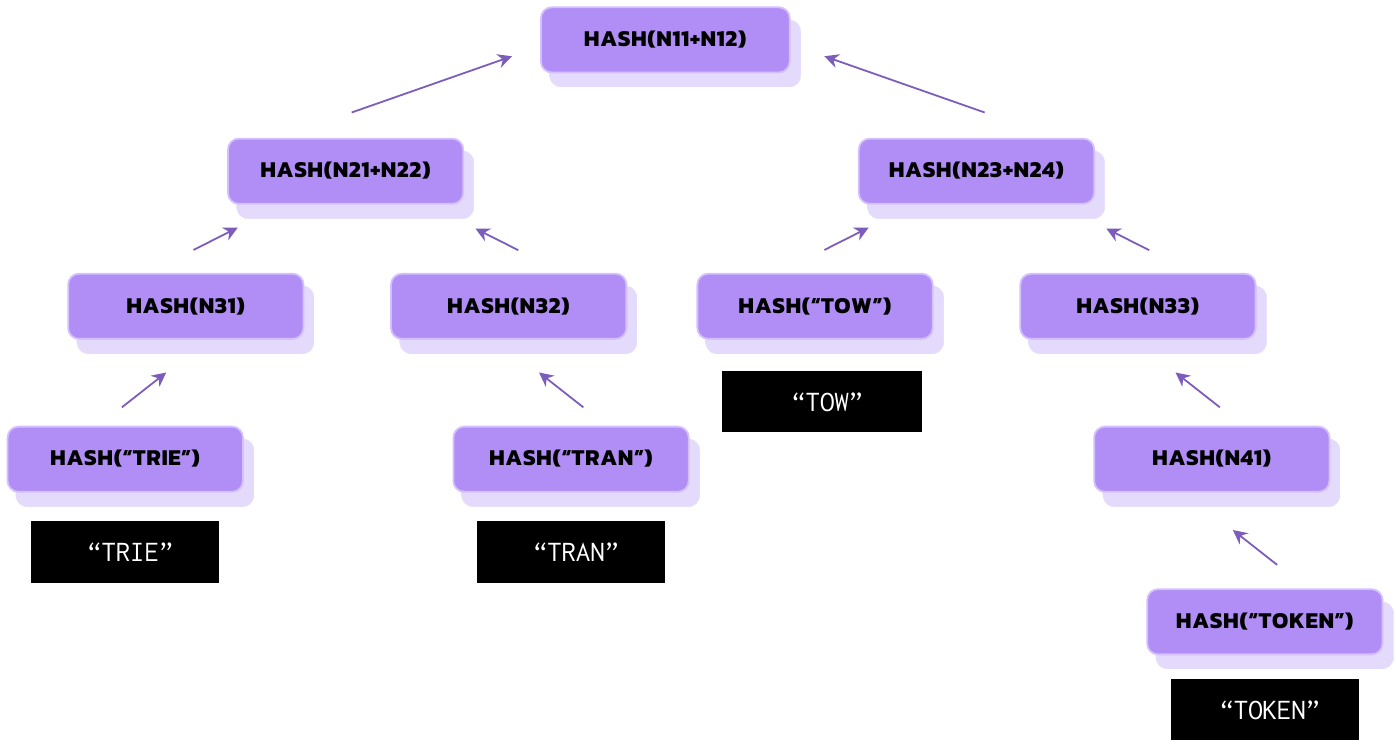}
    \caption{The Merkle-Patricia trie}
    \label{fig:merkle-patricia-trie}
\end{figure}

A Merkle-Patricia trie combines the features of both Merkle trees and Patricia tries, providing a cryptographically authenticated data structure that is space-efficient and allows for fast verification of data integrity and quick lookups. 
This hybrid structure is particularly prominent in the Ethereum blockchain, where it is used to manage the state of the system efficiently and securely.
Each trie node in Merkle-Patricia trie is compressed like in Patricia tries to optimize space.
Additionaly each node is hashed, and the hashes are used to secure the structure as in Merkle trees.
A single hash called root hash represents the entire structure, providing a secure fingerprint for verification.
The Merkle-Patricia trie is shown in the Fig.~\ref{fig:merkle-patricia-trie}

\subsection{Ethereum Merkle-Patricia tries}

\begin{figure}[ht]
    \centering
    \includegraphics[width = \textwidth]{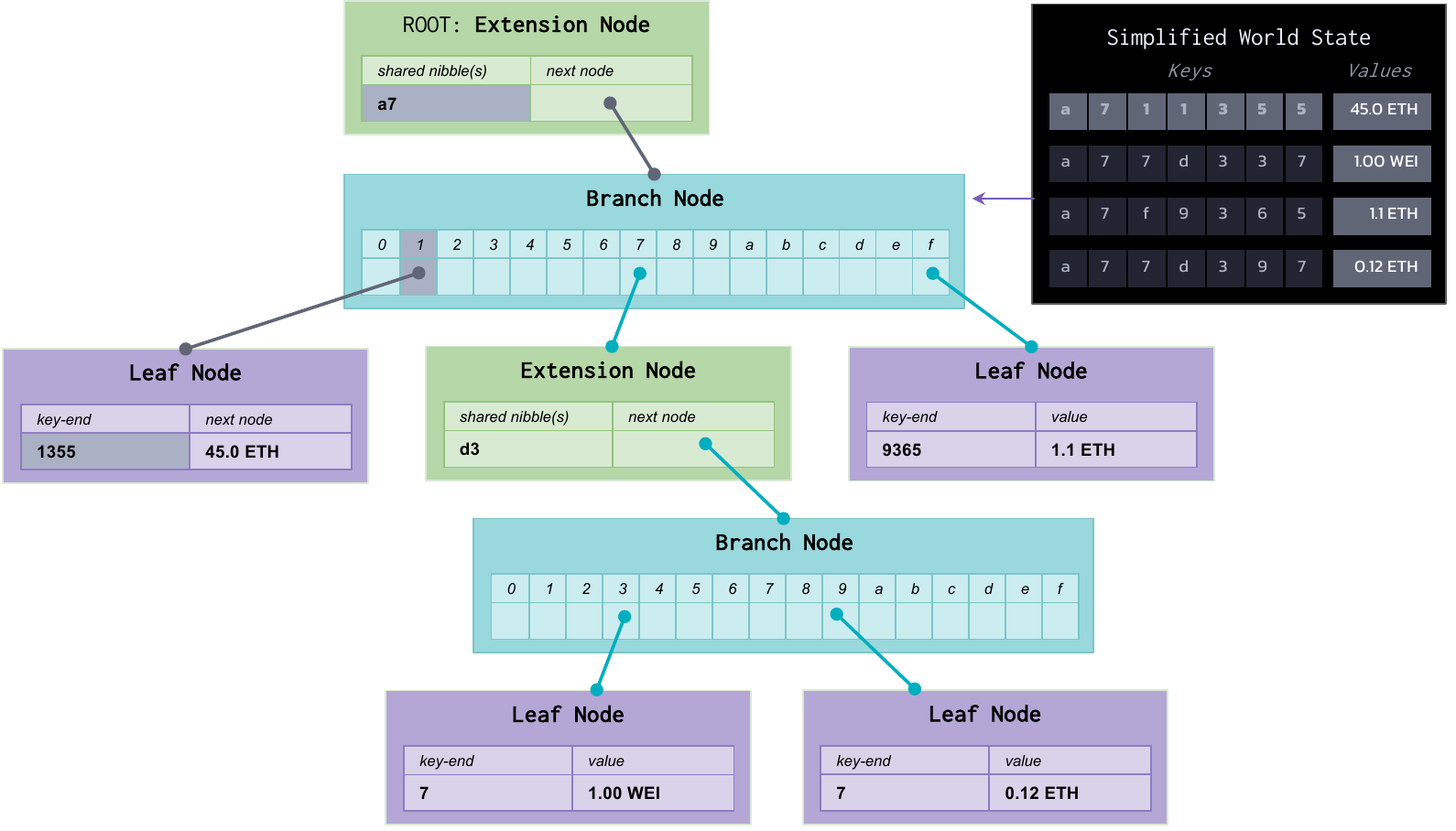}
    \caption{Ethereum Merkle-Patricia trie: Simplified Structure of Nodes and Key-Value Storage}
    \label{fig:eth-mpt}
\end{figure}

The Ethereum Merkle-Patricia trie (MPT) is a data structure used to manage and verify the state of the Ethereum blockchain. It uses a concept of Merkle-Patricia trie to provide a secure, compact, and efficient way to store and update key-value pairs.
The Ethereum MPT consists of several types of nodes:
\begin{itemize}
    \item \textbf{Root Node:} This is the entry point of the trie. It can be an extension node, branch node, or leaf node, depending on the structure of the data. In the Fig.~\ref{fig:eth-mpt} the Root Node is an Extension Node

    \item \textbf{Branch Node:} This node has 16 possible children, each corresponding to a hexadecimal character (0-9 and a-f). It can store a value if the path ends at this node. If a branch node does not have a child for a particular nibble, that child is null.

    \item \textbf{Extension Node:} This node is used to store shared nibbles (a nibble is half a byte, or four bits). If multiple keys share a common prefix, an extension node is used to reduce redundancy. It points to another node in the trie.

    \item \textbf{Leaf Node:} This node contains the end of a key and its corresponding value. Leaf nodes are terminal nodes, meaning they do not point to any other nodes.
\end{itemize}
\noindent
In the Fig.~\ref{fig:eth-mpt} a simplified view of an Ethereum Merkle-Patricia trie is presented.

\subsection{Versionised data structures}
The Merkle-Patricia trie facilitates the efficient storage and retrieval of historical states in Ethereum.
By combining the features of Merkle trees and Patricia tries, this structure allows for immutable and versioned data storage.
By keeping track of root hashes over time, different versions of the tree can be accessed. 
Each root hash corresponds to a specific state of the tree at a given point in time.
In Ethereum, the state of the blockchain at any block can be represented by a root hash of a Merkle-Patricia trie.
Accessing historical states involves referencing the root hash associated with a specific block.
\begin{figure}[ht!]
    \centering
    \includegraphics[width=\textwidth]{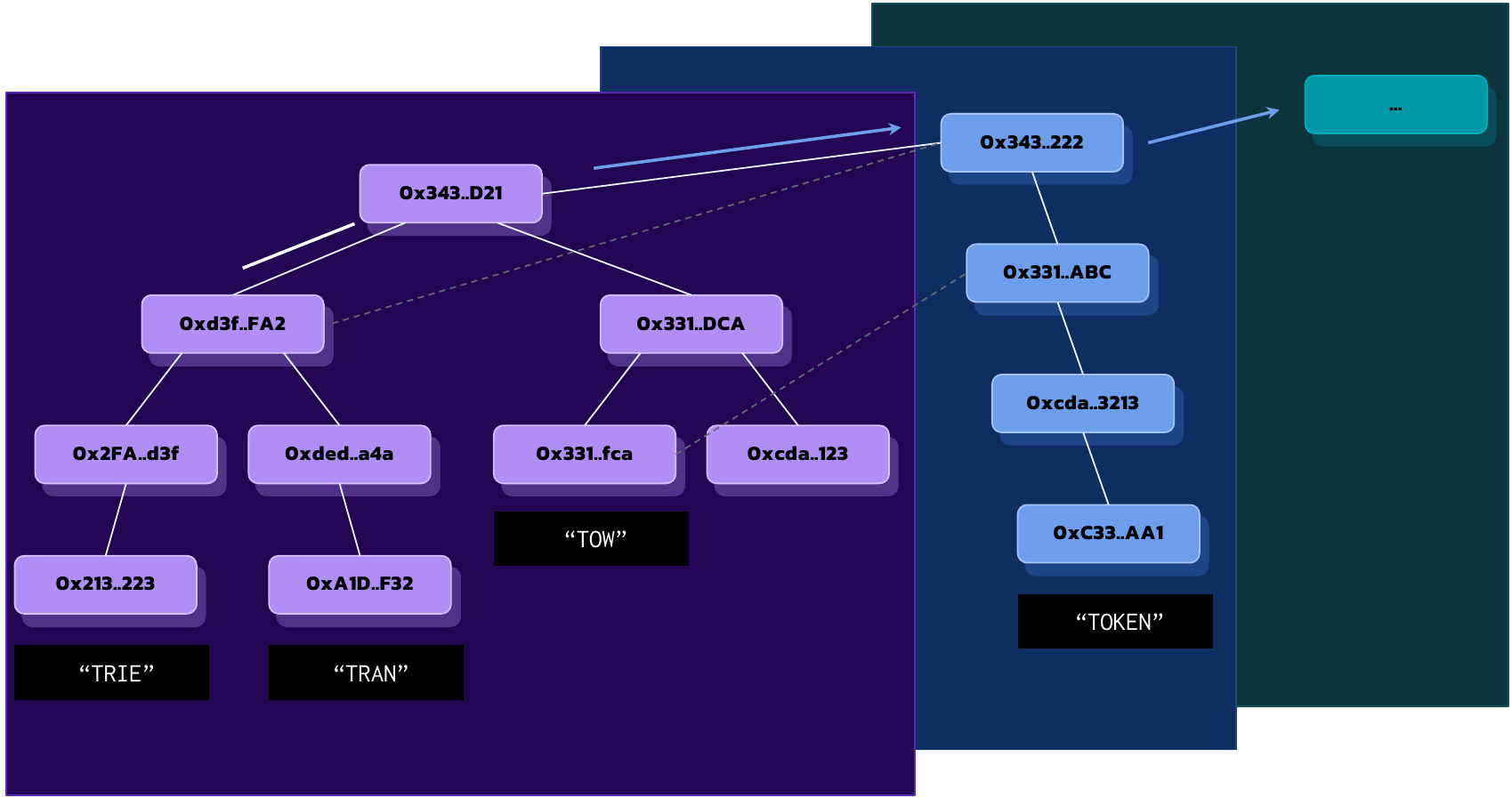}
    \caption{Merkle-Patricia trie as a versionised data structure}
    \label{fig:verionised-data-structure}
\end{figure}
As presented in Fig.~\ref{fig:verionised-data-structure} when a new  state change occurs, only the affected nodes and paths in the tree are updated, while unchanged parts of the tree are reused.
This efficient updating mechanism ensures that each historical state can be precisely reconstructed by referring to its corresponding root hash.
Additionally, the immutable nature of the tree, where each update results in a new tree structure without altering the previous ones, ensures that all historical states are preserved.
This enables Ethereum to provide proofs of inclusion and exclusion for any given state, leveraging the cryptographic properties of the Merkle tree.
Consequently, Merkle-Patricia trees enable secure, efficient, and verifiable access to historical data.

\subsection{Ethereum as a data structure}
Ethereum, as a decentralized platform for executing smart contracts, relies on a complex data structure to ensure the corectness of its operations. 
At the core of Ethereum structure there are three trie-based structures: the state trie, the transactions trie, and the receipts trie. 
These tries are integral to how Ethereum maintains and verifies the blockchain’s state, transactions, and logs, respectively. 
Each block in the Ethereum blockchain contains the root hashes of these tries, summarizing the state of the entire system at that block.
Ethereum uses Merkle-Patricia tries (MPT) for these key data structures as it combines the features of a Merkle tree (a way to verify the integrity and consistency of data) and a Patricia trie (making data retrieval efficient).
There are three key Trie Roots in Ethereum

\begin{itemize}
    \item State Trie (along with Storage Trie) 
    \item Transactions Trie
    \item Receipts Trie
\end{itemize}

Note that in Shapella update~\cite{ShapellaUpdate}, which included EIP-4895~\cite{EIP4895}, the Withdrawals Trie has been introduced.
Withdrawals are represented as a new type of object in the execution payload – an “operation” – that separates the withdrawals feature from user-level transactions.

\begin{figure}[ht!]
    \centering
    \includegraphics[width = \textwidth]{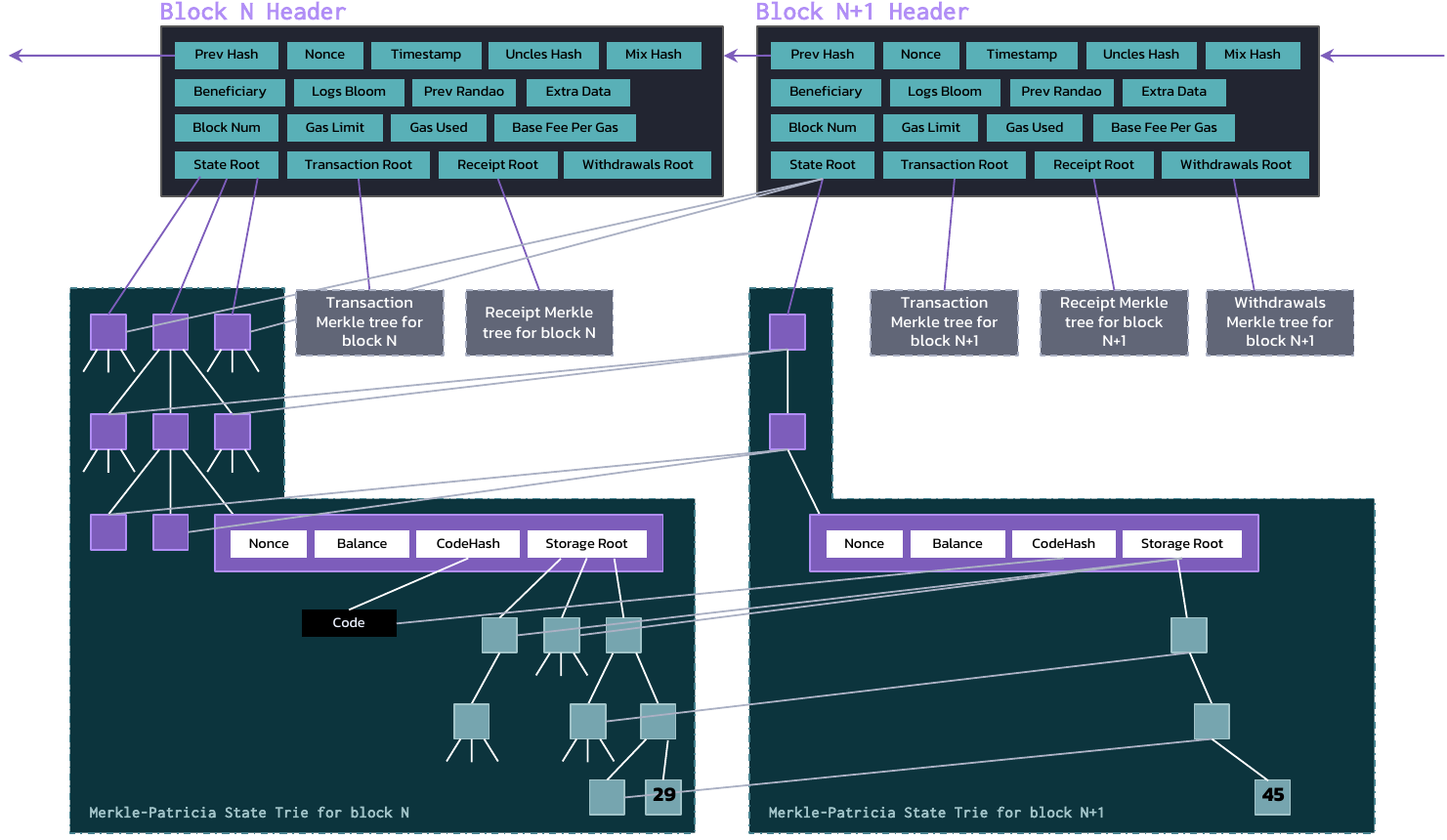}
    \caption{Illustration of Ethereum block structure. The diagram shows how each block header contains root hashes for state, transactions and recipts tries.}
    \label{fig:EthDataStructure}
\end{figure}

The state trie is a core component of Ethereum, storing the state of all accounts, both externally owned accounts (EOAs) and contract accounts. Each account has associated information, including nonce, balance, storage root, and code hash.
The state trie maps account addresses to account states. Each account state is 4-tuple -- nonce, balance, code hash and storage root. 
Storage root creates another trie if the account is a contract with its own storage.
The root hash of the state trie is stored in each block header. 
This root hash represents the entire state of the Ethereum network at the time of that block, allowing verification of any account's state with a Merkle proof.
The transactions trie stores all the transactions included in a block. Each transaction contains details such as sender, recipient, value, data, and gas information.
The transactions trie organizes transactions by their index in the block. 
Each node in the trie represents a transaction, allowing efficient verification and retrieval.
The root hash of the transactions trie is included in the block header. 
This hash provides a compact and verifiable summary of all transactions in the block, ensuring the integrity and order of transactions.
The receipts trie contains the receipts for all transactions in a block. 
A transaction receipt includes information about the transaction’s execution, such as the status (success or failure), gas used, and logs generated by events within the transaction.
Similar to the transactions trie, the receipts trie is indexed by the position of each transaction in the block.
Each receipt provides a detailed record of what happened during the execution of the transaction.
The root hash of the receipts trie is also stored in the block header. 
This root hash allows verification of transaction outcomes and associated logs, communicating with external world.
By storing these root hashes in the block header, Ethereum ensures that any changes in the blockchain’s state, transactions, or receipts can be efficiently detected and verified.

\section{Storage Proofs}
We identify storage proofs as critical in Ethereum for verifying the integrity and existence of data without requiring access to the entire dataset. 
This section explores the mechanisms and challenges associated with storage proofs in Ethereum. 
We begin with base storage proofs, which form the foundation for verifying individual pieces of data within the blockchain. 
We then examine the hierarchical structure of Ethereum's data to understand how storage proofs relate to state and header.
Next, we address the algorithm for proving historical state using Merkle Mountain Range and Merkle-Patricia trie constructions, examining their respective advantages and limitations. 
This is followed by a discussion on Keccak-256 performance, the hashing algorithm used in Ethereum, and its implications for the efficiency of zero-knowledge proofs. 
Finally, we discuss proving multichain storage proofs, which involves verifying data across Ethereum and its Layer 2 solutions. 

\subsection{Basic storage proof}

Proof of state and storage in the context of Ethereum, this involves proving that a specific state (account balance, nonce, code hash) or storage value (specific key-value pair in contract storage) is correctly included in the state or storage trie without the need to locally persist the entire trie structure.
Ethereum uses Merkle-Patricia tries for both the state trie and storage trie. 
State Trie mapps adresses to account states and storage trie maps storage keys to storage values for each contract.
\begin{figure}[ht]
    \centering
    \includegraphics[width=\textwidth]{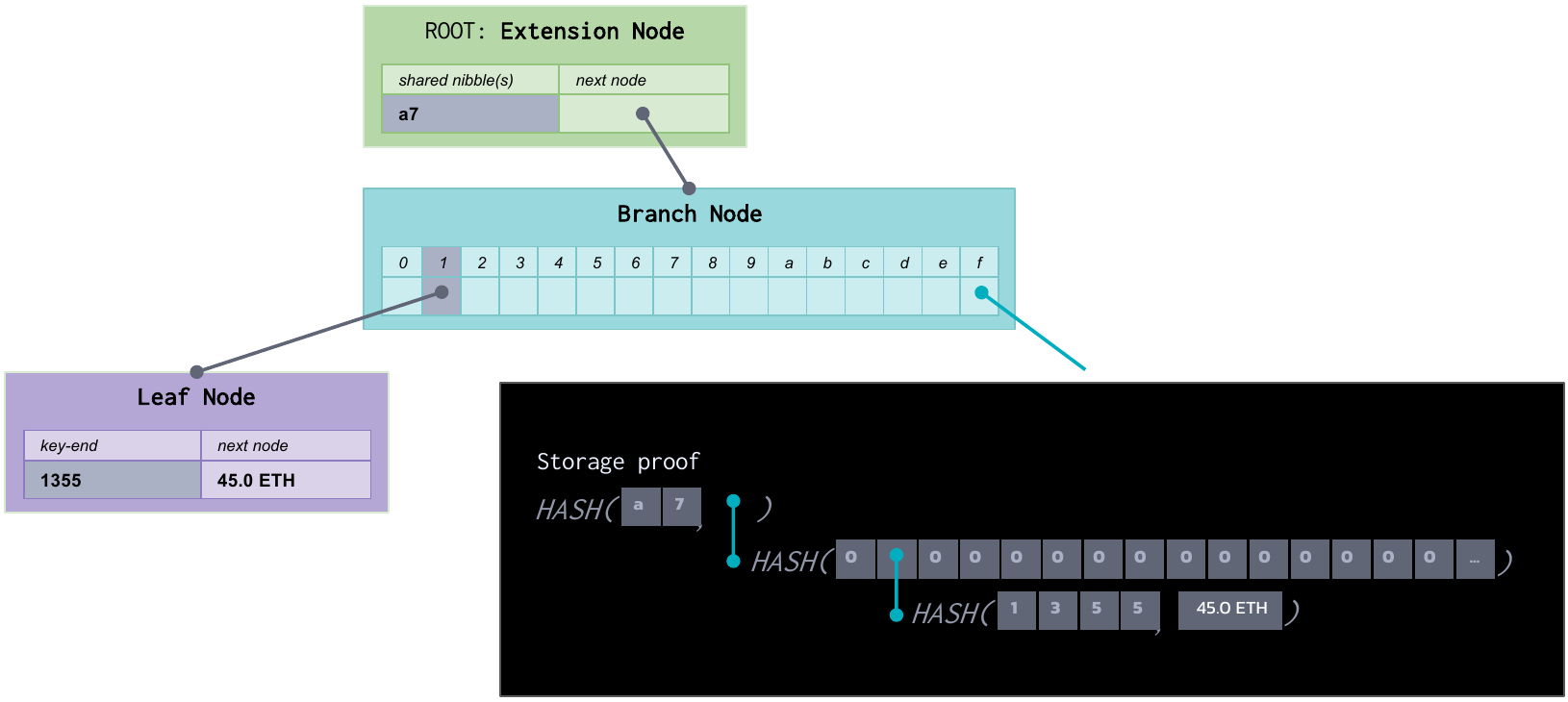}
    \caption{Storage proof in Ethereum Merkle-Patricia trie}
    \label{fig:eth-storage-proof}
\end{figure}
\noindent
To demonstrate that a specific leaf node (account or storage slot) is a part of the trie we construct a standard Merkle proof for the state or storage trie. 
The Fig~\ref{fig:eth-storage-proof} illustrates the structure of storage proofs in Ethereum using the Merkle-Patricia trie. 
It shows the hierarchical organization of nodes, including the root, branch, and leaf nodes, and how they are interconnected through shared nibbles and hash functions.

In the context of Ethereum, each storage proof involves multiple nodes, leading to large and quickly growing proof sizes. 
This is due to the necessity of including every node in the path from the root to the leaf in the proof. As the blockchain expands, these proofs become increasingly cumbersome, impacting efficiency.
To address this issue, compressing the proofs into zero-knowledge (ZK) circuits is emerging as a future approach.
ZK circuits can compress the Merkle proofs by enabling the verification of data without revealing the data itself, thus significantly reducing the size of storage proofs.
The root and leaf hashes are compressed as public inputs to the ZK-circuit with the path (Merkle proof) as the private ZK-circuit input. 
This approach significantly reduces the size of the proofs, making them much smaller.

\subsection{Hierarchy of proofs}

Storage proofs in Ethereum follow a hierarchical structure built around the \texttt{Header-State-Storage} relationship. 
Due to this hierarchical organization, proving the validity of a storage value requires a sequence of proofs. 
To prove what is inside a smart contract a basic storage proof has to be completed.
Then, another Merkle proof is needed, demonstrating that the storage root belongs to the state root. 
Final Merkle proof shows the state root belongs to the block header. 
This sequence forms a complete Ethereum storage proof, leveraging the inherent structure of Ethereum's data hierarchy to provide cryptographic verification of storage values.

\subsection{Historical state proof}
\begin{figure}[ht!]
    \centering
    \includegraphics[width=\linewidth]{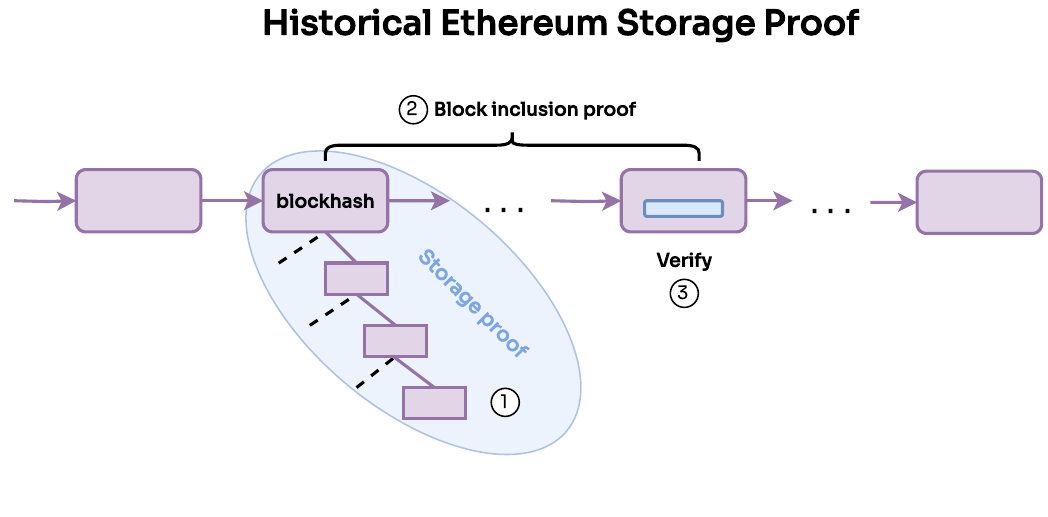}
    \caption{Diagram of the historical Ethereum Storage Proof}
    \label{fig:historical_storage_proof}
\end{figure}

To prove the historical data on Ethereum, one has to demonstrate that a particular part of state (eg. account balance, nonce, code hash, or storage value) existed at a specific block in the past.
Before examining historical state verification, we establish notation to differentiate between blocks at various depths in the chain:

\begin{itemize}
    \item \texttt{current\_block} - the block currently being mined
    \item \texttt{recent\_block} - a block not older than 256 blocks from \texttt{current\_block}
    \item \texttt{historical\_block} - a block older than \texttt{recent\_block} (i.e. older than 256 blocks from \texttt{current\_block})
\end{itemize}
\noindent
This distinction is important as Ethereum provides different mechanisms for accessing block hashes depending on their age. 
To verify a proof from \texttt{historical\_block} on the \texttt{recent\_block} we need to prove that \texttt{historical\_block} belongs to the same chain as \texttt{recent\_block}.
We call such construction a \textbf{block inclusion proof}.
As the historical state proof is verified against historical block hash (see Section 3.2 Hierarchy of proofs) we would like to build a block inclusion proof using block hashes. 
The process is illustrated in Fig.~\ref{fig:historical_storage_proof}.

For \texttt{recent\_block}, block hashes can be directly accessed using Solidity's built-in \texttt{blockhash()} function.
The \texttt{blockhash()} function can retrieve the block hash of one of the most recent 256 blocks, which simplifies the proof for these recent blocks.
However, accessing block hashes for \texttt{historical\_block} requires additional proving architecture, which we explore in the following sections.

To address this limitation, EIP-2935 proposes an extension of historical block hash accessibility~\cite{EIP2935}. 
It introduces a system contract that stores the last 8192 block hashes in a ring buffer structure. 
This mechanism allows for efficient retrieval of a broader range of historical block hashes directly from the state, enabling the creation of proofs for blocks beyond the 256-block limit without altering the existing \texttt{blockhash()} functionality. 
EIP-2935 is expected to be implemented as part of the Prague/Electra (Pectra) update, scheduled for delivery in Q4 2024 or Q1 2025. 
However, it's important to note that this solution still won't solve the problem for blocks older than 8192 block hashes away. 
Given Ethereum's average block time of about 12 seconds, 8192 blocks represent approximately 27 hours of blockchain history.

Naively, the problem with creating block inclusion proof for \texttt{historical\_block} could be solved the same way as in the case of two subsequent blocks.
We could do that by calculating a hash of the \texttt{historical\_block}, and check if it equals the \texttt{prevHash} of the \texttt{recent\_block}.
Although for two blocks separated by many others we would need to repeat that procedure many times and verify the \texttt{prevHash} for each subsequent block pair.
The problem is we cannot cache for all pairs of (\texttt{historical\_block}, \texttt{recent\_block}), as it would need a lot recomputing -- for each new block we would need to reiterate over all past blocks.

An efficient solution to construct block inclusion proof for \texttt{historical\_block} requires several key properties. 
First, we need a data structure that maintains a set of pre-proved block hashes.
This can be conceptualized as a map of block hashes where each hash is recursively proven to be correct in relation to its predecessors.
An important operation on this structure is the addition of new elements, which must be provable for proper construction. 
When adding a new block hash, we can recursively combine the existing proof with the proof for the new element, maintaining the correctness chain. 
With a structure defined in this way -- combining pre-proven elements with a provable addition operation -- we can recursively prove that any element belongs to the chain.

A Merkle tree could serve this purpose, as it provides the properties for recursive proof construction. 
However, for improved efficiency, we explore an alternative solution by introducing a specialized hierarchical data structure -- the Merkle Mountain Range.

\subsubsection{Merkle Mountain Range}
A Merkle Mountain Range (MMR) is a cryptographic data structure that optimizes the concept of a Merkle tree for growth.
They have been used for historical block hash accumulators by Herodotus~\cite{herodotus2024mmr}. 
An MMR consists of multiple Merkle trees, or so called “peaks,” that are structured in such a way that they form a range. 
Each peak in the range is a complete Merkle tree, and the number of peaks changes dynamically as new elements are added.
Unlike traditional Merkle trees, where adding a new element may require reorganizing the tree, MMRs allow for efficient appending of new elements without requiring an update of existing nodes.
That makes Merkle Mountain Ranges well designed to handle grow only data sets.
Hence, MMRs are immutable structures, meaning once a peak or any other node is created, it cannot be altered.
This helps with efficiency as with new element there is no need to re calculate all of the nodes.
The proving scheme in the MMR is similar to the one in ordinary Merkle tree.
As there is no single arbitrary root in MMR, we create it by hashing together all of the peaks.
Thus, Merkle proof in Merkle Mountain Range consists of standard Merkle proof and list of the peaks. 

\begin{figure}[ht!]
    \centering
    \includegraphics[width=\linewidth]{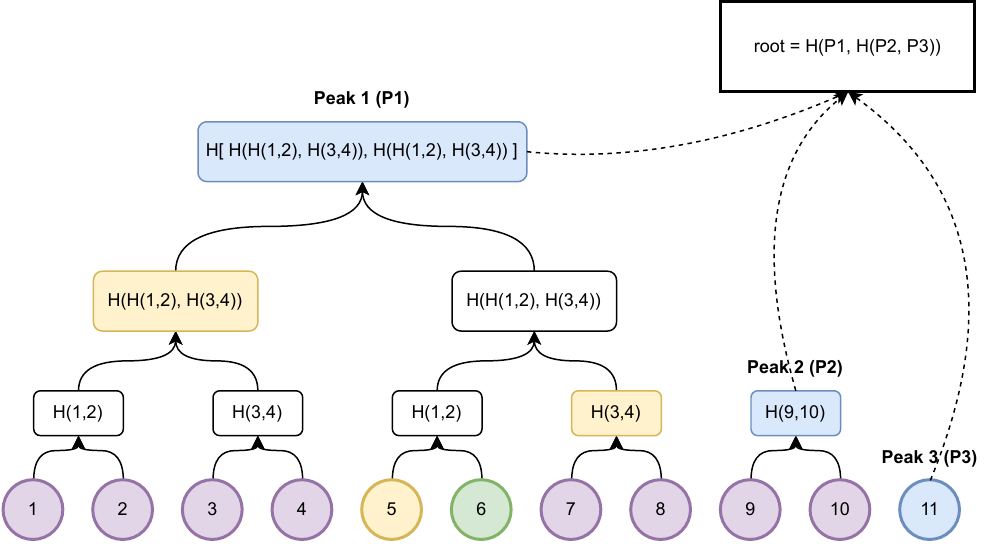}
    \caption{Proving scheme in the MMR.}
    \label{fig:mmr-proving-scheme}
\end{figure}

\noindent
In the Fig~\ref{fig:mmr-proving-scheme} a proving scheme of element 6 (green) in the 11-element MMR is presented.
Yellow blocks constitute a standard Merkle proof of leaf 6 belonging to its peak (Peak 1).
Blue nodes represent peaks.
Blue and yellow nodes hashed together form a Merkle proof of element 6 belonging to the MMR.
Please note that the root is not an element of the MMR.

\subsubsection{Block inclusion proof with MMR}
As described in Section 3.4, an efficient proving chain inclusion between \texttt{recent\_block} and \texttt{historical\_block} requires a more complex approach than direct block hash verification. 
An efficient solution can be achieved by building a Merkle Mountain Range structure on top of the block hash list. The proving system consists of three steps presented below.

\vspace{10pt}
\noindent\fbox{
    \parbox{\textwidth}{
\noindent    
\begin{enumerate}
    \item Initialize MMR with a single block
    \item For the MMR with a single block and  calculate ZK-proof $\pi$ of proper construction 
    \item For each new block in the chain calculate the recursive proof $\pi$ and update Merkle Mountain Range. 
    By appending new element to the MMR we need to
    \begin{itemize}
        \item Update ZK-proof $\pi$
        \item Update MMR nodes
        \item Update MMR root by hashing updated peaks
    \end{itemize}
\end{enumerate}
}
}
\vspace{10pt}

\begin{figure}[ht]
    \centering
    \includegraphics[width=\textwidth]{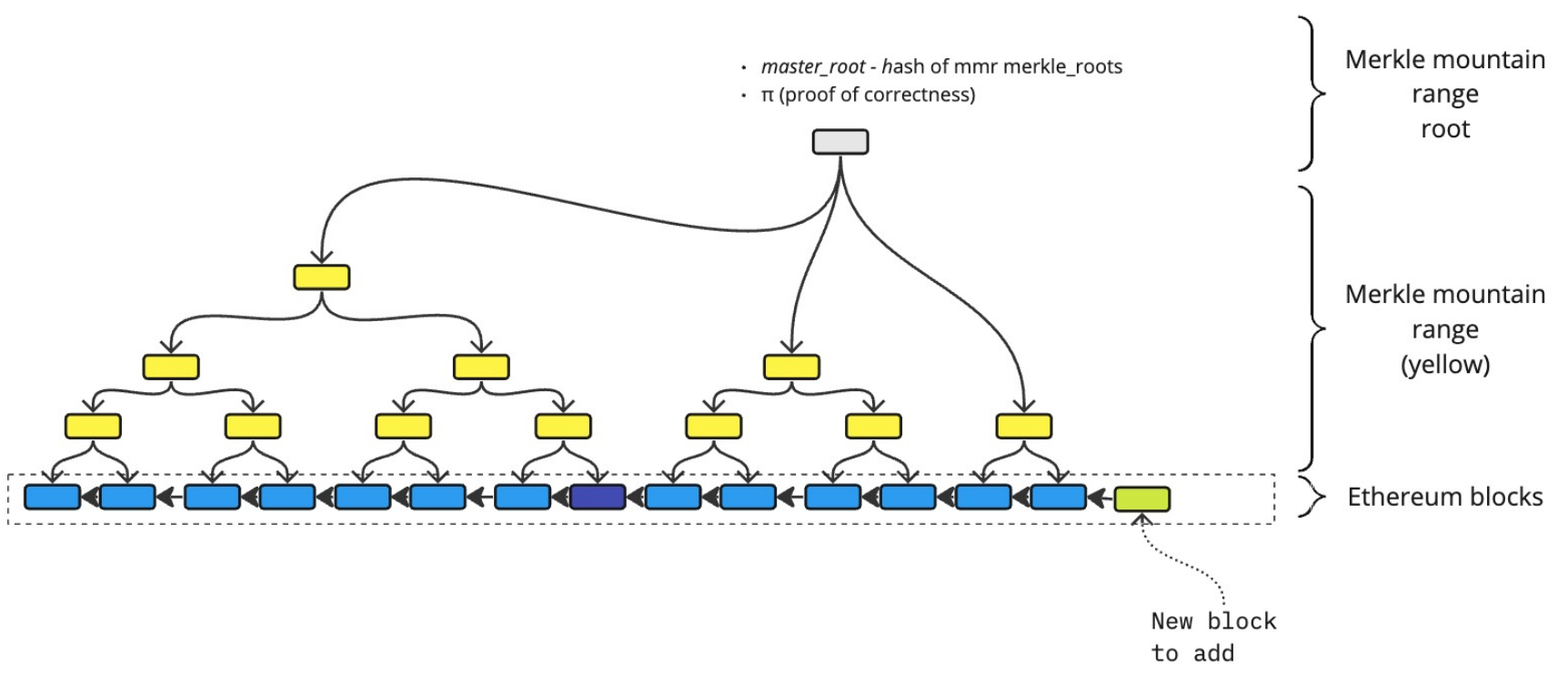}
    \caption{Diagram illustrating the structure of a Merkle Mountain Range (MMR) used to efficiently cache the block hashes for inclusion proofs needed in historical state proving.}
    \label{fig:MMR}
\end{figure}

\noindent
The proof of proper construction $\pi$ is a zero-knowledge (ZK) proof. 
This is an important aspect of the algorithm because zero-knowledge proofs allow one party to prove to another that a statement is true without revealing any information beyond the validity of the statement itself. 
In the context of MMR, $\pi$ ensures that the structure of the MMR and the incremental proofs are constructed correctly without revealing the underlying data of each block.
The whole process is illustrated in the Fig.~\ref{fig:MMR}

The Merkle Mountain Range (MMR) approach, while offering efficiency advantages for recent block hash proofs due to its peak structure, faces limitations when dealing with very old block hashes. 
As the time depth increases, the efficiency gains diminish, with the root calculation becoming as computationally intensive as a standard Merkle tree.
Furthermore, the MMR's inherent single-direction growth restricts its flexibility, particularly in scenarios requiring prepending of elements.
To address these limitations, we propose an alternative solution based on Ethereum's native Merkle-Patricia trie (MPT) data structure.

\subsubsection{Block inclusion proof with MPT}

To address the limitations of the MMR approach and provide a comprehensive solution for block inclusion proofs, we propose an adaptation of the Merkle-Patricia trie (MPT) data structure, similar to Ethereum's native implementation. 

The proposed MPT implementation differs from Ethereum's native MPT by storing \texttt{<Block Number, blockhash>} tuples in the trie nodes, as opposed to original \texttt{<Key, Value>} tuples in Ethereum structure. 
This modification allows for a direct mapping of block numbers to their corresponding block hashes.
This design choice of using MPT for \texttt{<Block Number, blockhash>} pairs enables efficient and flexible manipulation of the cached data, particularly for extending the sequence in both directions -- by appending and prepending new blocks.
Similarly to the MMR attempt, alongside the MPT structure, there must exist a ZK-proof $\pi$ of proper construction that verifies the correct addition of new elements through append and prepend operations.
The append and prepend invariants which are verified to assure proper construction are presented in the box below

\vspace{10pt}
\noindent\fbox{
    \parbox{\textwidth}{
\noindent
For the append operation, which adds a new rightmost block to the sequence, the following condition must be satisfied:
\begin{equation}
\forall (i, h_i) \in T 
~\exists (i+1, h_{i+1}) \in T:     \mathcal{B}_A\texttt{.prev\_hash} = h_i \wedge \texttt{HASH}(\mathcal{B}) = h_{i+1},
\end{equation}
\noindent
where $T$ is the Merkle-Patricia trie used for caching block hashes, $\mathcal{B}_A$ is the block being appended, $h_i$ represents block hash of the current rightmost block, while $h_{i+1}$ is the block hash of the bock being appended. 

\vspace{10pt}
\noindent
For the prepend operation, which adds a new leftmost block, the following condition must be met:
\begin{equation}
    \forall (i, h_i) \in T 
    ~\exists (i-1, h_{i-1}) \in T:  \mathcal{B}_{LM}\texttt{.prev\_hash} = h_{i-1} \wedge \texttt{HASH}(\mathcal{B}) = h_{i},  
\end{equation}
\noindent
where $\mathcal{B}_{LM}$ is the current leftmost block and $h_{i-1}$ represents the block hash of the block being prepended, while the rest of the symbols retain their meaning from equation (1).
}}
\vspace{10pt}

Whether we choose MMR or MPT for our block inclusion proofs, we need zero-knowledge proof $\pi$ to verify proper construction of these structures.
This presents an interesting challenge: Ethereum uses Keccak-256 for all its hash operations by default, but Keccak is not particularly efficient in zero-knowledge contexts. 
For more efficient implementation of the proposed block caches, we should consider building them with alternative hash functions that are more zero-knowledge friendly. 
The next section explores this performance challenge and describes several promising alternatives.

\subsection{Keccak-256 performance challenge} 

Zero-knowledge proofs used to verify data integrity must compute plenty of hash functions. 
However, Keccak-256, the hash function used throughout Ethereum, presents significant challenges in zero-knowledge contexts. 
Its internal structure relies on complex bitwise operations and multiple permutation rounds, making it particularly inefficient in SNARK circuits.
This inefficiency has driven the development of alternative hash functions specifically designed for zero-knowledge applications. 
These ZK-friendly hash functions prioritize algebraic operations over bitwise operations, significantly reducing computational overhead in ZK circuits.

Several promising alternatives have emerged in recent years. 
Poseidon and Starkad share a common HadesMiMC structure, optimized for different environments - Poseidon for prime fields and Starkad for binary fields~\cite{Grassi2019Starkad}. 
Blake3 takes a different approach, achieving efficiency through parallelization~\cite{BLAKE3}. 
MiMC opts for simplicity, using basic algebraic structures to achieve both security and performance~\cite{cryptoeprint:2016/492}.
Other designs include Rescue, which combines the familiar sponge construction with ZK-specific optimizations~\cite{cryptoeprint:2019/426}.
The Pedersen hash function, while based on elliptic curve operations, has found practical application in privacy-focused systems through optimizations developed for the Zcash protocol~\cite{ZCash}.

Each of these alternatives offers different trade-offs between security, efficiency, and implementation complexity. 
For the block caching structures presented in the Section 3.3, they present an opportunity to significantly improve proof generation performance compared to Keccak-256.

\subsection{Multichain state proof}

The Ethereum ecosystem has evolved to include numerous Layer 2 (L2) solutions, built to enhance scalability while leveraging Ethereum's security. 
L2 networks operate on top of the Ethereum mainnet and handle transactions off-chain, significantly reducing the load on the mainnet.
Even though L2 chains work independently of L1 they use Ethereum as a source of security.
There is a variety of L2 chains on Ethereum based on one of three solutions: ZK Rollup~\cite{ethereum_zk_rollups}, Optimistic Rollup~\cite{ethereum_optimistic_rollups, optimism_rollup_overview} or Based Rollup~\cite{ethresearch_based_rollups, taiko_based_rollup_faq_2023}. 
This evolution has led to fragmentation across different L2 networks, each processing transactions independently. 
Cross-chain interactions between these networks remain complex and slow, as data retrieval and transaction verification require asynchronous communication through Ethereum mainnet.

\begin{figure}[ht!]
    \centering
    \includegraphics[width=\textwidth]{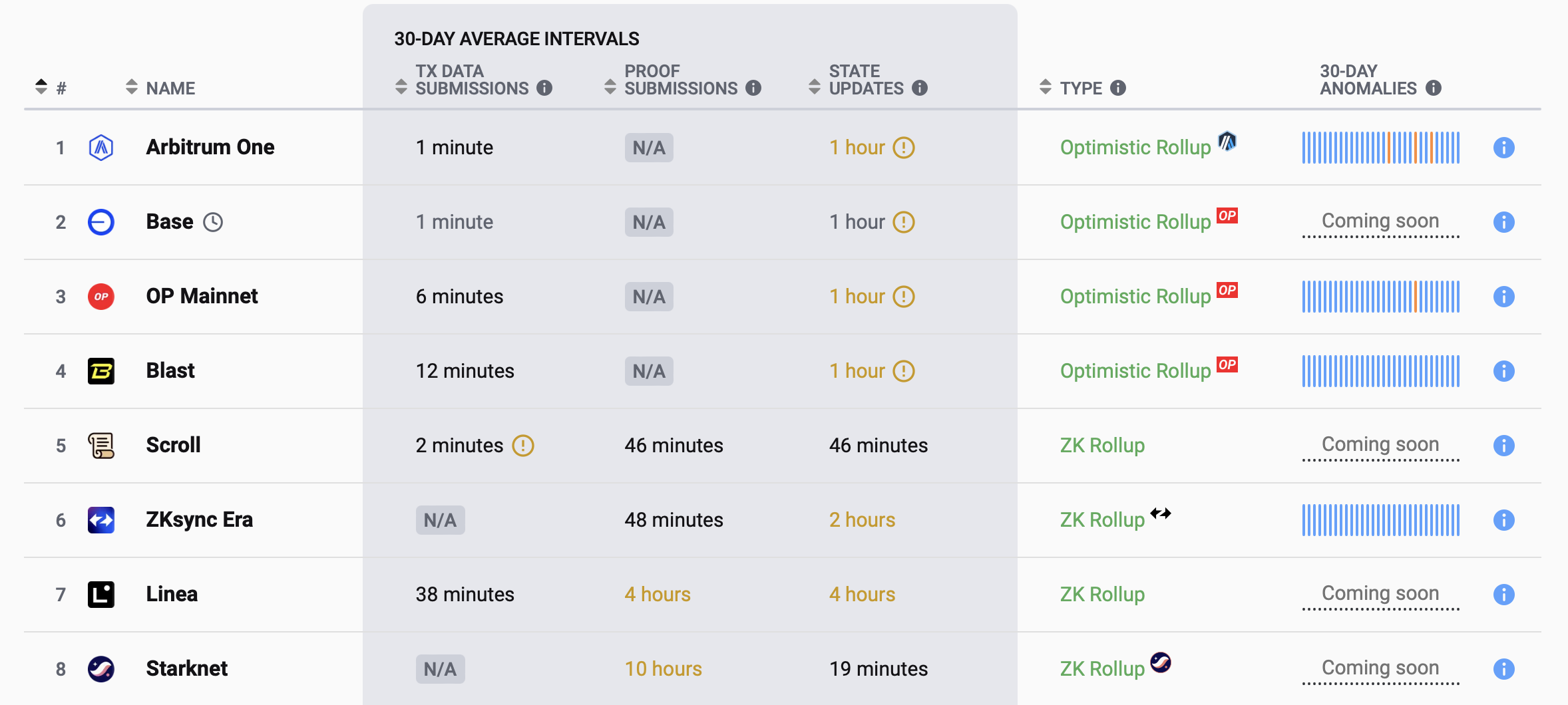}
    \caption{Comparison of L2 networks showing average state update intervals over 30 days. Courtesy of \texttt{L2beat.com}}
    \label{fig:L2-state-updates}
\end{figure}

Multichain storage proofs offer an elegant solution to this challenge.
Instead of relying on complex bridging mechanisms, they enable direct verification of state across different chains. 
Before exploring specific verification patterns, we need to understand how L2 networks maintain their security through periodic state updates and proof submissions to Ethereum.
In the Fig~\ref{fig:L2-state-updates} a table is shown comparing various L2 networks based on their 30-day average intervals for transaction data submissions, proof submissions, and state updates. 
State updates in L2 Ethereum networks involve changes to the information that the network holds about accounts, balances, and smart contract states.

Essentially, every rollup is protected by submitting its state and transactions to the mainchain.
L2 periodically submits its blockhash to L1 for further verification.
Thanks to that we can verify L2 state by examining L1 block hash.
That lets us to introduce multichain proofs -- a proofs between different L2 chains and Ethereum.
Due to the location of Verifier contract and source of the Storage Proof, we distinguish three cases of multichain proofs:

\begin{enumerate}
    \item Verification of Storage Proof from L1 on L2
    \item Verification of Storage Proof from L2 on L1
    \item Verification of Storage Proof from one L2 verified on another L2
\end{enumerate}

\subsubsection{Finality in multichain proofs}

\begin{figure}[ht!]
    \centering
    \includegraphics[width=\textwidth]{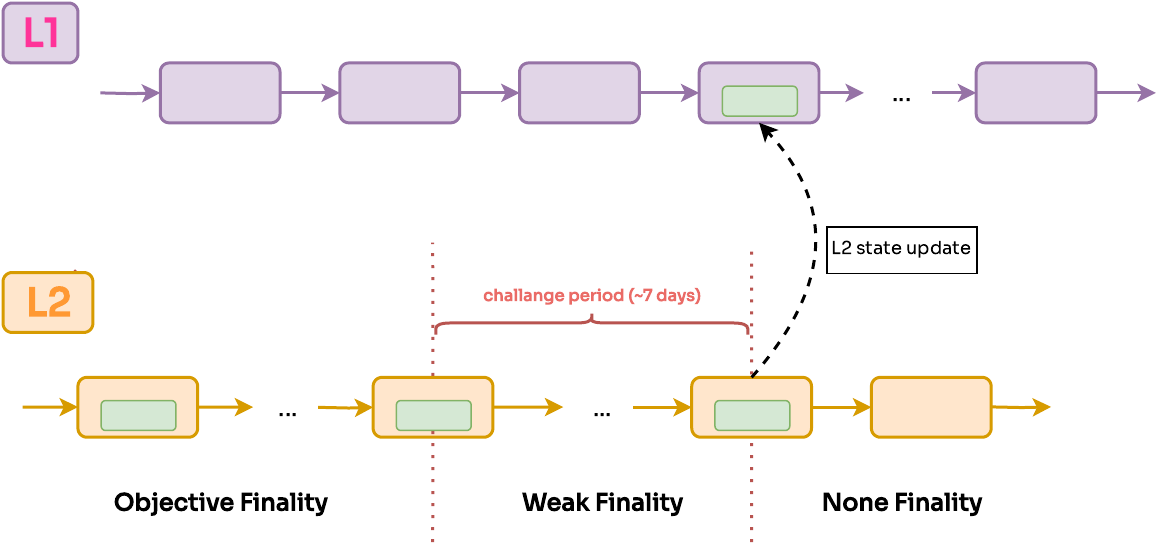}
    \caption{Finalities on optimistic rollup L2. Some state (denoted by green rectangle inside the L2 block) has None Finality until L2 state update happens. Once state update happens Weak Finality is achieved which transforms into Objective Finality after challenge period.}
    \label{fig:l2-finality}
\end{figure}

Prior to discussing the multichain proving and verifying architectures we need to discuss an aspect of the proving system which is the finality on both L1 and L2.
On Ethereum, finality is typically achieved within around 13 minutes. 
This duration aligns with the time required to generate a storage proof and verify that a block belongs to L1. Consequently, the block header with a storage proof submitted to L2 occurs only after L1 finality is ensured.
The status of finality on L2 is more complicated.
As illustrated in the Figure~\ref{fig:l2-finality}, we distinguish L2 three types of finality which another open three cases

\begin{itemize}
    \item None Finality: This state occurs for blocks that have not sent a state update to L1. 
    These blocks are not finalized and are subject to changes.
    \item Weak Finality: Blocks that have sent a state update to L2 but are within the challenge period fall into this category. 
    These blocks can be contested using fraud proofs.
    It takes 30-60 mins to reach a weak finality.
    \item Objective Finality: Once the challenge period passes without successful fraud proofs, blocks achieve strong finality. 
    This status assures that the block is correct and immutable.
    Is is accomplished after around 7 days.
\end{itemize}

The timing characteristics of these finality states have important implications for cross-chain verification. 
State update latency manifests in the transition from None Finality to Weak Finality, typically occurring within 30-60 minutes, depending on the L2 network's batch submission frequency.
This is important for protocol designer considerations.
For example in a protocol, during this period, proofs might not be considered valid for high-value transactions.
The seven-day challenge period for Optimistic Rollups creates a trade-off between security and finality speed. 
This way, an exemplary protocol could require wait for Objective Finality for high-value cross-chain transactions unless additional L2 security mechanisms are in place. 
Applications may choose to await Objective Finality for maximum security, accept Weak Finality with additional application-level security measures, or implement a hybrid approach with escalating confidence levels based on time elapsed since state update.

Depending on when the L1 block header with the storage proof is sent to L2, verification might be performed on a block that is not finalized (none finality) or is weakly finalized (in a challenge period).
Note that achieving weak finality on L2 takes longer than finality on L1 because state updates on L2 occur approximately every 30 minutes to 1 hour.

\subsubsection{Verification of Storage Proof from L2 on L1}

\begin{figure}[ht!]
    \centering
    \includegraphics[width=\textwidth]{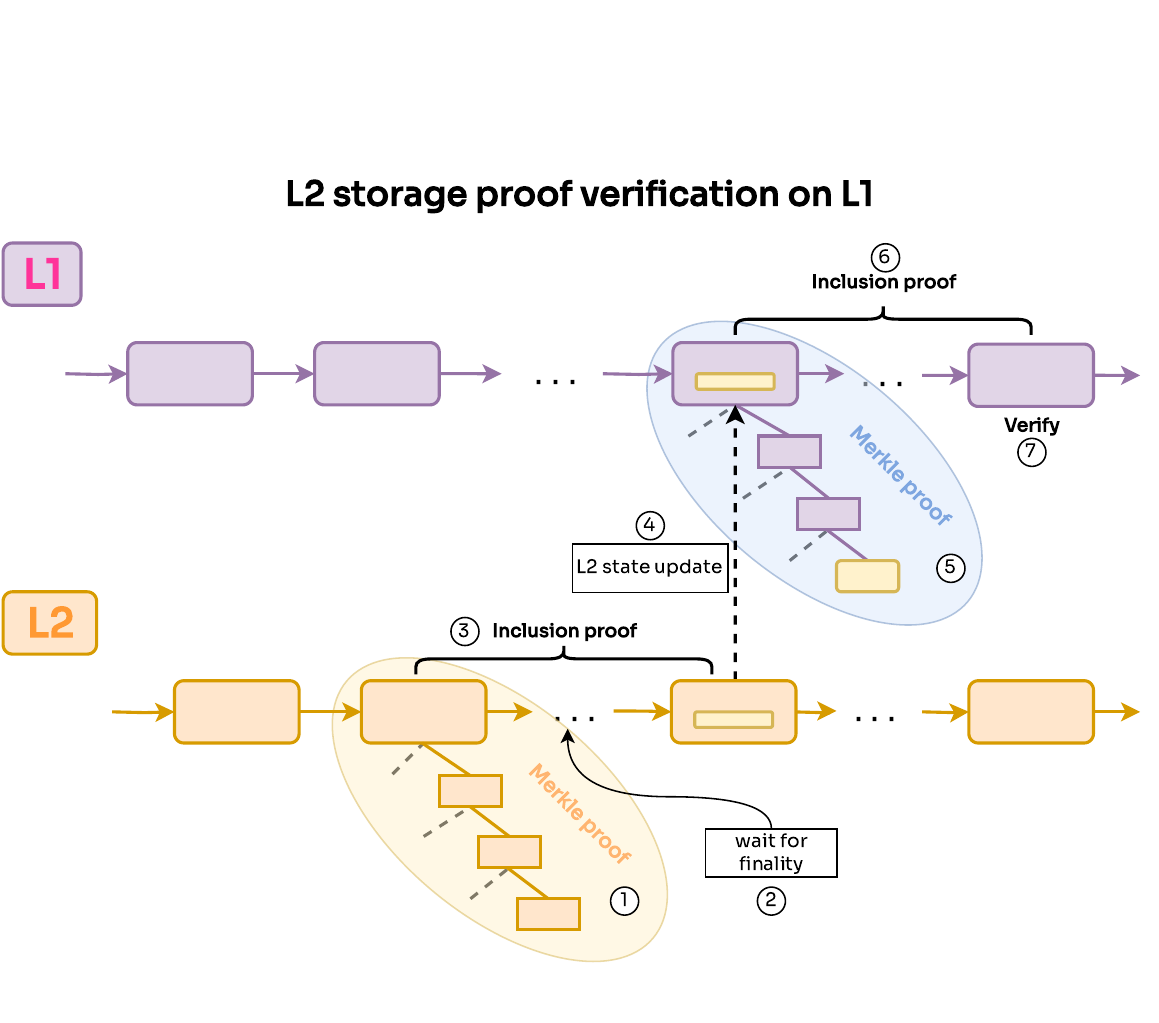}
    \caption{Diagram of multichain proving system where L2 storage proof is verified on L1.}
    \label{fig:L2toL1Diagram}
\end{figure}

We start with the case where storage proof in created on Layer 2 and verified on Ethereum.
First the proving system and data structure on L2 has to be consistent with one used at Ethereum.
That's why we focus on optimistic rollups like Optimism or Arbitrum, where the storage proofs can be created the same way as on L1.
The process of verification of storage proof from L2 on L1 is shown in Fig.~\ref{fig:L2toL1Diagram}.

\noindent
To verify storage proof from L2 on L1, we follow these steps:

\begin{enumerate}
    \item A Merkle proof of the storage is created at block on L2. That block is called \texttt{L2proofBlock}
    \item At chain dependent time intervals (see Fig.~\ref{fig:L2-state-updates}), L2 blockhash is transfered to L1 with state update mechanism. 
    This ensures weak finality for the block called \texttt{L2transferedBlock}.
    \item To assure that \texttt{L2proofBlock} belongs to the same chain as \mbox{\texttt{L2transferedBlock}} a block inclusion proof is needed.
    \item With state update mechanism \texttt{L2transferedBlock} lands on L1 block which is called \texttt{L1proofBlock}
    \item A storage proof of \texttt{L2transferedBlock} block hash belonging to \texttt{L1proofBlock} is created on L1
    \item Before verification an inclusion proof is needed to verify that \texttt{L1proofBlock} and \texttt{L1verificationBlock} both belong to the same chain.
    \item The proof is verified on the \texttt{L1verificationBlock} 
\end{enumerate}

\subsubsection{Verification of Storage Proof from L1 on L2}

\begin{figure}[ht!]
    \centering
    \includegraphics[width=\textwidth]{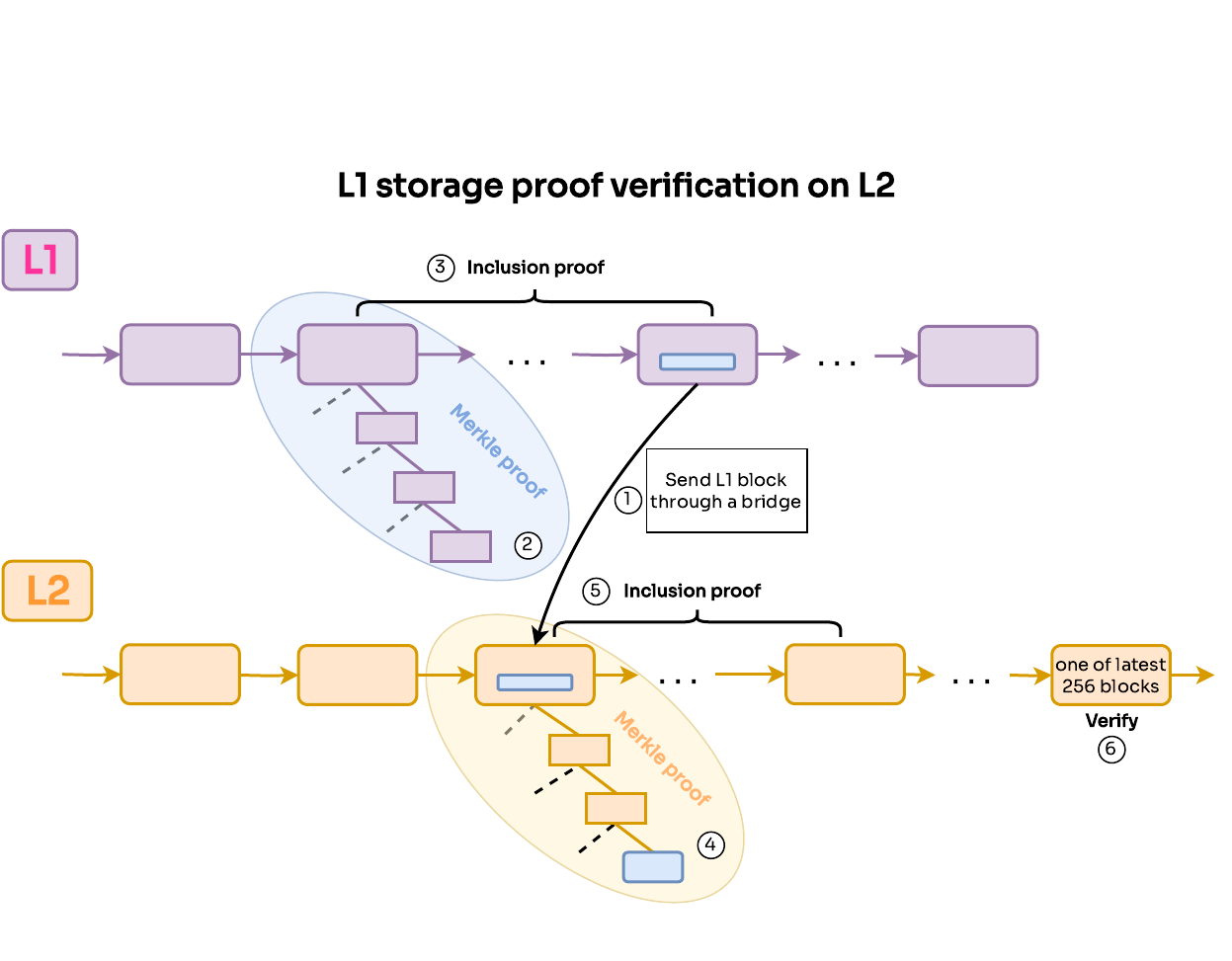}
    \caption{Diagram of multichain proving system where storage proof from L1 is verified on L2.}
    \label{fig:L1toL2Diagram}
\end{figure}

The verification of storage proofs from Ethereum mainnet (L1) on Layer 2 networks presents a distinct set of challenges and requirements compared to $\text{L2}\rightarrow\text{L1}$ verification. 
This verification pattern must address the fundamental asymmetry in the relationship between L1 and L2 networks, where L2s maintain continuous awareness of L1 state through bridge mechanisms. 
Unlike $\text{L2}\rightarrow\text{L1}$  verification, where state updates provide a natural pathway for proof verification, $\text{L1}\rightarrow\text{L2}$  verification requires careful consideration of block hash transmission and synchronization between layers. 
To verify storage proofs from L1 on L2, we propose a verification architecture that leverages the existing bridge mechanisms for block hash transmission. 
This capability, exemplified by the \texttt{L1blockHash()} function in Optimism's OP stack, is not universally available across L2 implementations. 
A schematic representation of verification of storage proof from L1 on L2 is presented in the Figure~\ref{fig:L1toL2Diagram}

\noindent
To verify storage proof from L1 on L2, we follow these steps:

\begin{enumerate}
    \item First of all L1 block hash is transferred via secure bridge to L2.
    We refer to that block as \texttt{L1transferedBlock}.
    \item A Merkle proof of the storage is created at block on L1. 
    We call this block \texttt{L1proofBlock}
    \item To assure that \texttt{L1proofBlock} belongs to the same chain as \texttt{L1transferedBlock} we need to verify an inclusion proof.
    Note that we are only able to verify the claims for the blocks from the last transfer and older.
    \item The L2 block where \texttt{L1transferedBlock} lands is refered to as \texttt{L2proofBlock}.
    \mbox{A~storage} proof of \texttt{L1transferedBlock} block hash belonging to \texttt{L2proofBlock} is created on L2
    \item We run block inclusion proof on L2 to verify that \texttt{L2proofBlock} belongs to the same chain as \texttt{L2verificationBlock}
    \item That proof is verified on \texttt{L2verificationBlock}
\end{enumerate}

\subsubsection{Verification of Storage Proof from one L2 verified on another L2}

The verification of storage proofs between two Layer 2 networks can be formalized as a composition of $\text{L2}\rightarrow\text{L1}$ and $\text{L1}\rightarrow\text{L2}$ verification processes. 
This composition requires specific compatibility conditions between the participating L2 chains and their respective interactions with the Layer 1 network.

For secure cross-L2 verification, we need to ensure finality of block hashes on both L1 and the destination L2.
This means waiting for the source L2's state update to be finalized on L1, and then for that L1 block hash to be securely transmitted to the destination L2. 
It's worth noting that block hash transmission between L1 and destination L2 happens on the settlement layer regardless of the verification status, providing a foundation for the verification process.

\begin{figure}[ht!]
    \centering
    \includegraphics[width=0.85\textwidth]{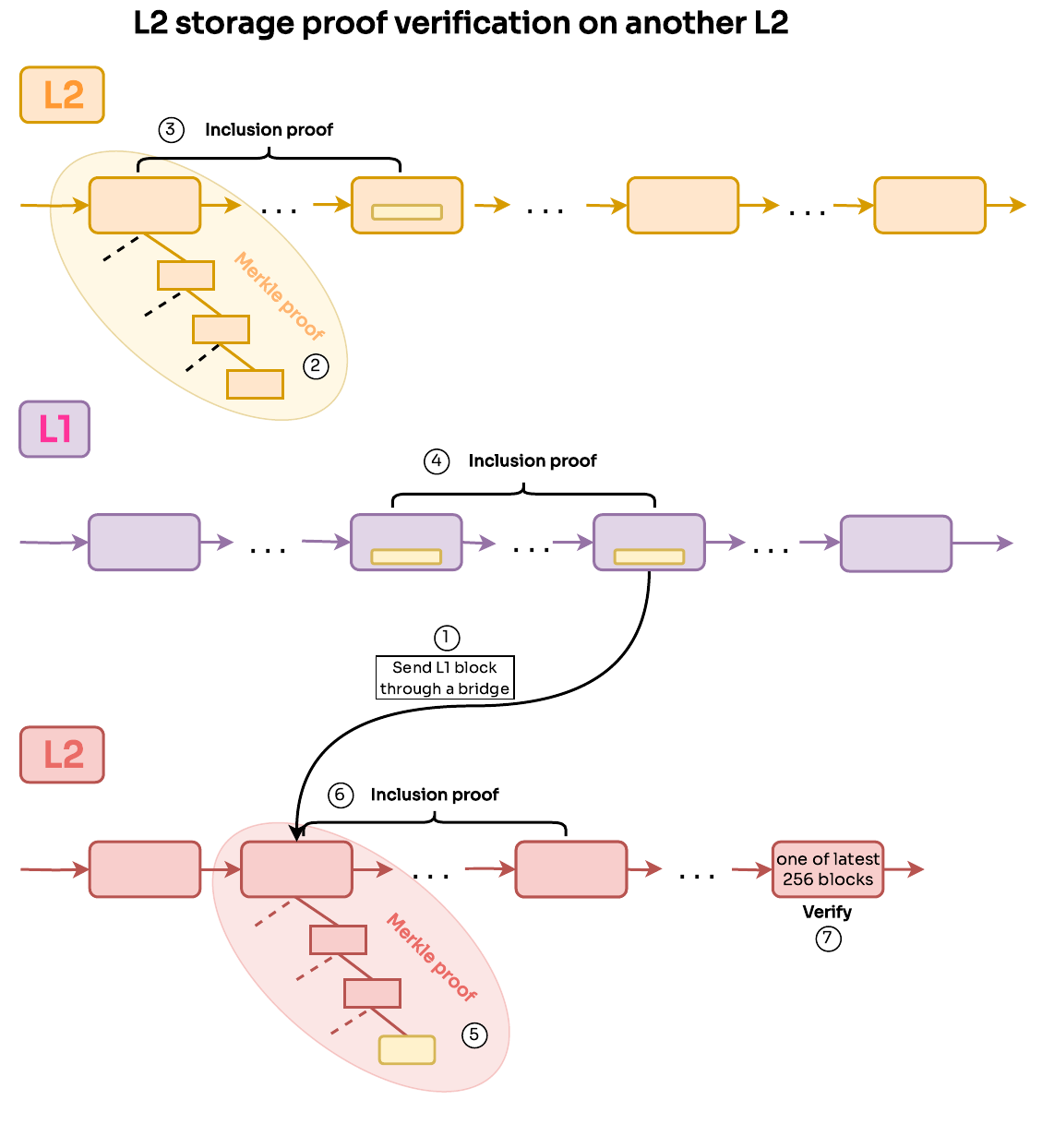}
    \caption{Diagram of multichain proving system where storage proof from L2 is verified on another L2.}
    \label{fig:L2toL2Diagram}
\end{figure}

A schematic representation of verification of storage proof from L2 on another L2 is presented in the Figure~\ref{fig:L2toL2Diagram}.
As described in Section 3.6.3 we start with transferring a L1 block hash via secure bridge to the destination L2.
We refer to that block as \texttt{L1transferedBlock}.
For the source L2 network, as described in Section 3.6.2, the verification pathway begins with creating a Merkle proof at \texttt{sourceL2ProofBlock},followed by awaiting the state update mechanism to transfer block hash to L1. 
The source L2 block hash is then transferred to L1, accompanied by an inclusion proof verifying that \texttt{sourceL2ProofBlock} belongs to the same chain as the block being transferred. 
Then, a block inclusion proof needs to be constructed at L1 to assure that block where L2 block hash landed and \texttt{L1transferedBlock} belong to the same chain.
The verification pathway on the destination L2 remains unchanged compared to section 3.5.3.
The destination L2 must create a storage proof of the L1 block hash belonging to \texttt{L2ProofBlock} and confirm chain membership through an inclusion proof.
Finally the proof is verified on the destination L2.

\section{Summary}
In this work we've been exploring the complexities and performance challenges associated with implementing storage proofs. 
While these proofs offer significant advancements, their complexity can deter developers and reduce the efficiency of smart contracts.
To address these challenges, the paper presents three cases of multichain storage proofs, designed to verify data across multiple interconnected blockchains, especially considering Ethereum and its Layer 2 solutions. 

Our analysis demonstrates that historical storage proofs can be effectively implemented using both Merkle Mountain Range and Merkle-Patricia trie structures, with MPT offering superior flexibility for bidirectional chain growth. 
We have shown that while EIP-2935 presents a partial solution for historical block access, a more comprehensive approach using MPT enables unlimited historical depth without sacrificing efficiency.
The investigation of multichain verification architectures reveals distinct patterns for L2$\rightarrow$L1, L1$\rightarrow$L2, and L2$\rightarrow$L2 proof verification. 
These patterns account for the asymmetric relationship between layers and varying finality characteristics across different networks. 
Described verification frameworks maintain security guarantees through careful consideration of state updates, bridge mechanisms, and inclusion proofs, establishing a foundation for reliable cross-chain state verification.
Performance challenges, particularly those related to Keccak-256 in zero-knowledge contexts, have been addressed through the analysis of alternative ZK-friendly hash functions. 
This analysis provides insights into the trade-offs between security, efficiency, and implementation complexity in cross-chain verification systems.

The architectures and methodologies presented in this paper present a theoretical and practical framework for implementing robust historical and multichain storage proofs in the Ethereum ecosystem. 
These review contribute to the broader development of scalable and interoperable blockchain systems.

\bibliography{refs}

\end{document}